\documentclass[usenatbib]{mnras}

\usepackage[T1]{fontenc}

\DeclareRobustCommand{\VAN}[3]{#2}
\let\VANthebibliography\thebibliography
\def\thebibliography{\DeclareRobustCommand{\VAN}[3]{##3}\VANthebibliography}

\usepackage{graphicx}	
\graphicspath{{Images/}}
\usepackage{amsmath}	
\usepackage{amssymb}	
\usepackage[table,xcdraw]{xcolor}    
\usepackage{subfig}
\usepackage{comment}
\usepackage[normalem]{ulem}
\usepackage{physics}
\usepackage{xcolor}

\definecolor{royalazure}{rgb}{0.0, 0.22, 0.66}
\definecolor{auburn}{rgb}{0.43, 0.21, 0.1}
\definecolor{bostonuniversityred}{rgb}{0.8, 0.0, 0.0}
\definecolor{planet}{HTML}{69DF45}
\newcommand{\GI}{\raisebox{-.15\height}[0pt][0pt]{\includegraphics[width=0.13in]{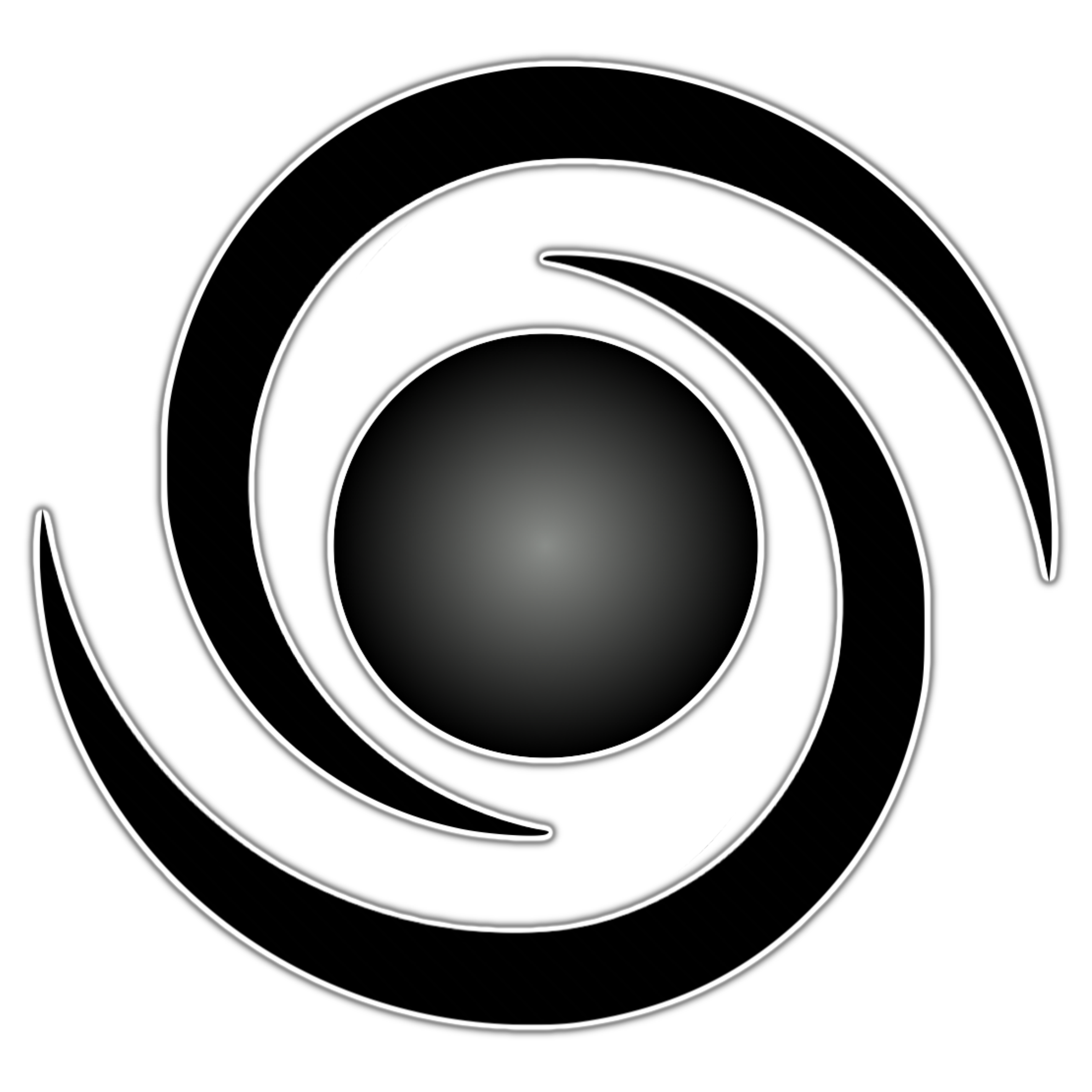}}}
\newcommand{\planet}{\raisebox{-.15\height}[0pt][0pt]{\includegraphics[width=0.13in]{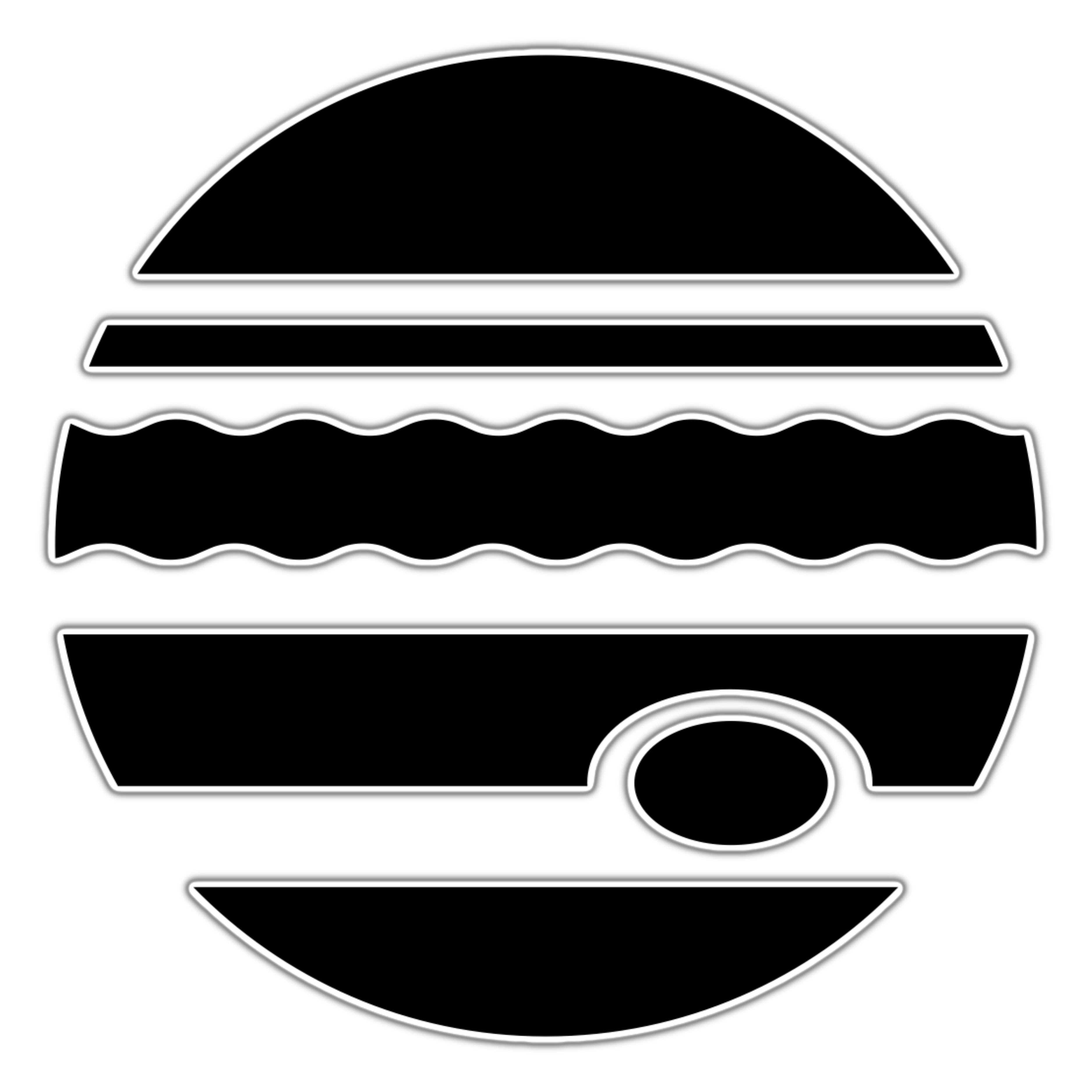}}}
\newcommand{\GIplanet}{\raisebox{-.27\height}[0pt][0pt]{\includegraphics[width=0.15in]{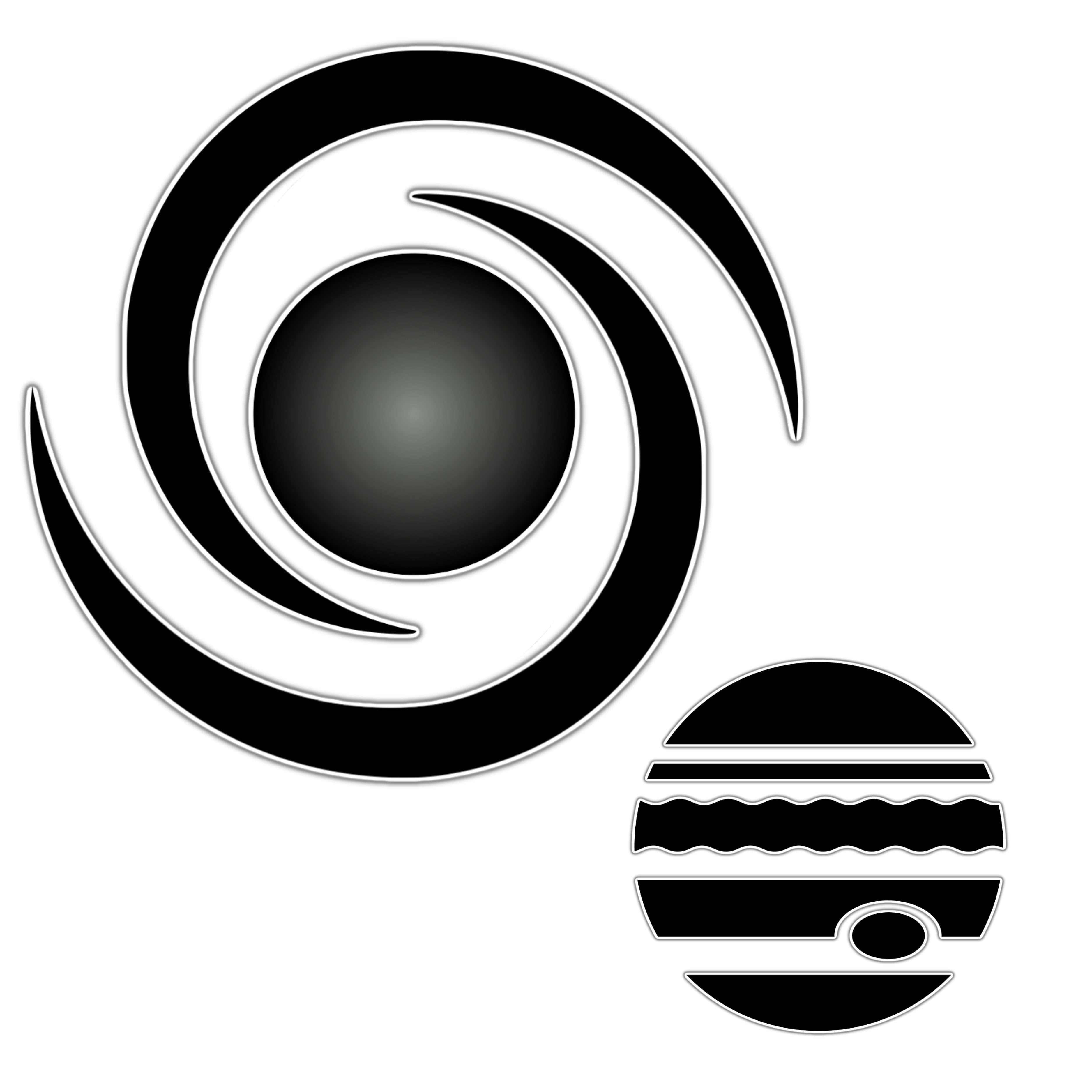}}}

\usepackage{hyperref}
\hypersetup{
    colorlinks=true,
    linkcolor=bostonuniversityred,
    linktoc=all,
    filecolor=blue,      
    urlcolor=royalazure,
    citecolor=royalazure
}

\usepackage{lipsum}

\usepackage{newtxtext,newtxmath}

\title[Planet-Disc Interactions in a GI disc]{Continuing to Hide Signatures of Gravitational Instability in Protoplanetary Discs with Planets }

\author[Rowther et al.]{
Sahl Rowther,$^{1, 2, 3}$\thanks{E-mail: sahl.rowther@leicester.ac.uk}
Rebecca Nealon,$^{1,2}$
Farzana Meru$^{1,2}$
\\
$^{1}$Centre for Exoplanets and Habitability, University of Warwick, Coventry CV4 7AL, UK\\
$^{2}$ Department of Physics, University of Warwick, Coventry CV4 7AL, UK\\
$^{3}$ School of Physics and Astronomy, University of Leicester, Leicester LE1 7RH, UK
}

\date{Accepted XXX. Received YYY; in original form ZZZ}

\pubyear{2022}

\begin{document}
\label{firstpage}
\pagerange{\pageref{firstpage}--\pageref{lastpage}}
\maketitle

\begin{abstract}
    We carry out three dimensional smoothed particle hydrodynamics simulations to study the impact of planet-disc interactions on a gravitationally unstable protoplanetary disc. We find that the impact of a planet on the disc's evolution can be described by three scenarios. If the planet is sufficiently massive, the spiral wakes generated by the planet dominate the evolution of the disc and gravitational instabilities are completely suppressed. If the planet's mass is too small, then gravitational instabilities are unaffected. If the planet's mass lies between these extremes, gravitational instabilities are weakened. We present mock Atacama Large Millimeter/submillimeter Array (ALMA) continuum observations showing that the observability of large-scale spiral structures is diminished or completely suppressed when the planet is massive enough to influence the disc's evolution. Our results show that massive discs that would be expected to be gravitationally unstable can appear axisymmetric in the presence of a planet. Thus, the absence of observed large-scale spiral structures alone is not enough to place upper limits on the disc's mass, which could have implications on observations of young Class I discs with rings \& gaps.
\end{abstract}




\section{Introduction}

In the last few years the Atacama
Large Millimeter/submillimeter Array (ALMA) has provided us with a large number of well resolved discs at millimeter wavelengths. These observations have revealed the ubiquity of disc substructures. The most common of these substructures being axisymmetric rings \& gaps.
\citep{2015ALMA,2016Andrews,2018Fedele,2018Andrews,2018Huang,2018Dipierro,2018Long,2020Booth}. A few of these discs are thought to be very young ($<1$Myr) \citep{2015ALMA,2018Fedele,2018Dipierro}. 
In their youth discs are expected to be more massive and in the regime where the disc self-gravity drives its evolution. In this phase of their lives, young massive discs develop gravitationally instabilities in the form of large-scale spiral structures. These discs are often referred to as GI discs. However, even young Class I discs show evidence of rings \& gaps \citep{Segura-Cox2020,2018Sheehan}.  Although rare, there is evidence that discs with spiral features in the midplane exist \citep{2016Perez, 2018Huang}. \textit{Is the rarity of spiral structures in the disc midplane due to young discs being less massive than we expect? Or can gravitational instabilities in massive discs be hidden by other processes?}

Protoplanetary discs are primarily made up of circumstellar gas H$_2$, which is difficult to observe and makes measuring the mass of the disc challenging. Thus, the total mass of the disc has to be inferred from the other constituents of the disc, the dust or molecular gas (usually CO). By assuming a globally constant dust-gas mass ratio, one can infer the gas mass from the more easily observed dust. However, the inferred gas mass remains very uncertain due to the assumptions that go into the disc radius, dust-to-gas mass ratio, dust opacities, distribution of grain sizes, or local over/under-densities \citep[and references therein]{2014Testi,Andrews_2015}. Inferring the disc mass using CO has other problems. The disc masses measured can be underestimated if the CO line emission is optically thick, and hence doesn't trace the disc midplane \citep{Bergin2017}. Additionally, measurements of the disc mass using CO can also be affected by freeze-out onto icy grains \citep{2001Zadelhoff} and photodissociation \citep{2015Reboussin}. However, \cite{2019Booth} showed that disc masses can be larger when using $^{13}$C$^{17}$O, an optically thin isotopologue which is a more robust tracer of the disc mass. In HL Tau, \cite{2020Booth} found that the disc mass estimates pushed the outer regions of the disc into the gravitationally unstable regime. Interestingly, the continuum data for HL Tau shows axisymmetric ring \& gap structure \citep{2015ALMA} instead of spiral structures; in contrast to what is expected from massive gravitationally unstable discs.

\begin{figure*}
    \begin{center}
    \includegraphics[width=\textwidth]{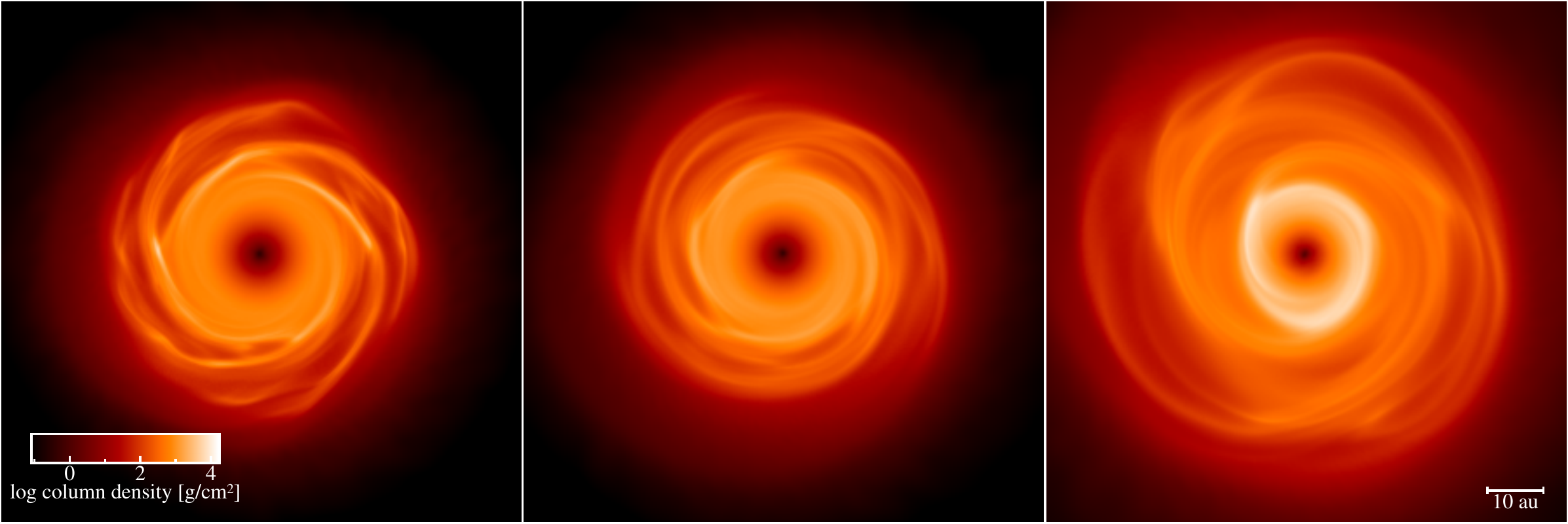}
    \caption[]{Surface density rendered plots of a $0.08M_{\odot}$ (left), $0.1M_{\odot}$ (middle), and a $0.25M_{\odot}$ (right) disc after 10 orbits at $R = R_\mathrm{out}$, right before a planet is embedded. Spiral structures due to gravitational instabilities are seen in all three discs to different extents. The higher the disc mass, the more gravitationally unstable the disc becomes as highlighted by the decreasing size of the gravitationally stable smooth inner disc as the disc mass increases.}
    \label{fig:InitialDisc}
    \end{center}
\end{figure*}

The physical interaction between a planet and the disc can alter the disc structure through the exchange of angular momentum \citep{2012Kley}. Thus it is natural to invoke planets as an explanation for the disc substructures such as rings \& gaps. Comparisons of hydrodynamical and radiative transfer simulations to observations from ALMA \citep{2018Clarke,2018Zhang} have validated this interpretation. The evidence for planet-disc interactions became stronger by \cite{2019Pinte,2020Pinte} where it was shown that the planet could be detected by the localised kink it produces in the gas kinematics. Given the young ages of some of these discs with rings \& gaps, if they were formed by planets then it's reasonable to assume the planet may have formed when the disc was younger and perhaps gravitationally unstable. Hence, it's necessary to understand how planet-disc interactions impact gravitationally unstable discs and their observability. 

Recent work has shown that the evolution of a gravitationally unstable disc can be altered by the same mechanisms that are often used in lower mass discs, as shown in \cite{2020bRowther} with a gap-opening planet, and by a warp in \cite{2022Rowther}. In the case of a planet, its spiral wake influences global properties of the disc. While in a warped disc, the non-coplanar geometry results in an oscillating radial pressure gradient, which alters the velocities in the disc.  In both, the disc was heated up rendering it gravitationally stable with an axisymmetric ring \& gap structure.

In this work we perform three-dimensional global numerical
simulations to extend the study in \cite{2020bRowther} by exploring a wider set of planet and disc masses to fully understand the interplay between planet-disc interactions and gravitational instabilities, and its implications on observations. This paper is organised as follows. In \S\ref{sec:model} we describe the simulations presented in this work. In \S\ref{sec:results} we present our results on how planet-disc interactions impact gravitationally unstable discs. The limitations and observational implications of this work are discussed in \S\ref{sec:disc}. We conclude our work in \S\ref{sec:conc}.

\section{Model}
\label{sec:model}

We use \textsc{Phantom}, a smoothed particle hydrodynamics (SPH) code developed by \cite{2018Price} to perform the suite of simulations presented here.

\begin{figure*}
    \begin{center}
    \includegraphics[width=\textwidth]{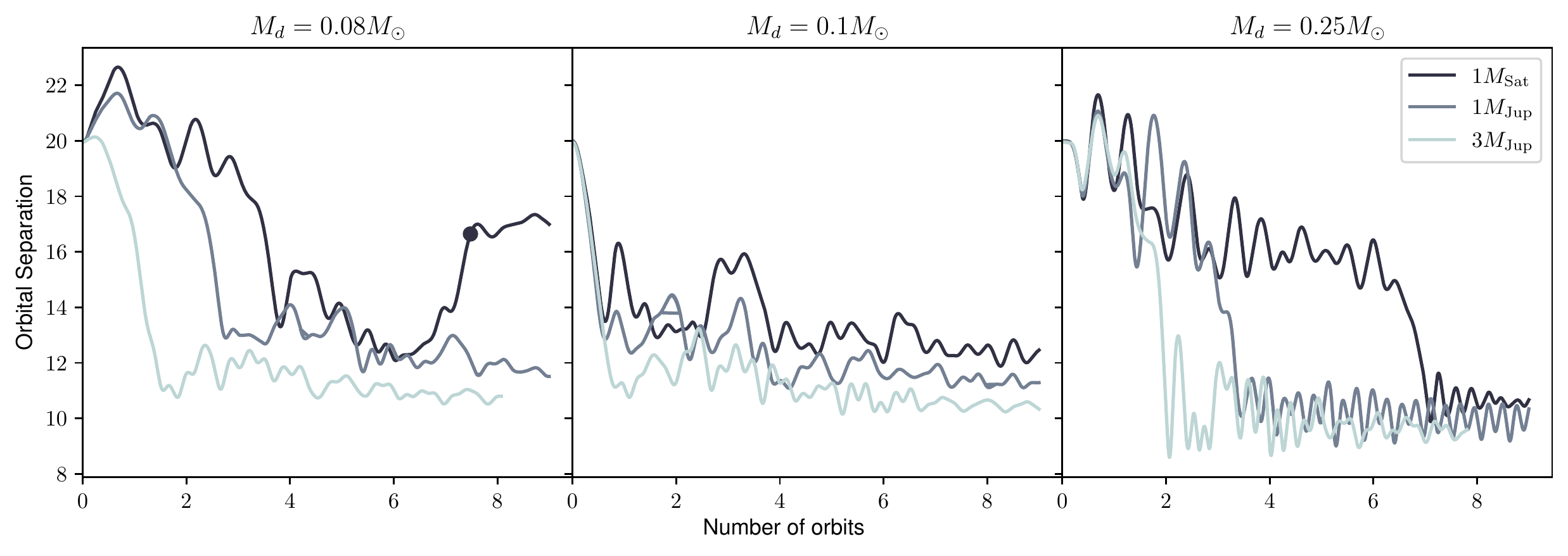}
    \caption[]{The migration tracks of a $1M_{\mathrm{Sat}}$, $1M_{\mathrm{Jup}}$, and a $3M_{\mathrm{Jup}}$ planet in a $0.08M_{\odot}$ (left), $0.1M_{\odot}$ (middle), and $0.25M_{\odot}$ (right) disc (as shown in Figure \ref{fig:InitialDisc}). The planets are able to slow their migration in the $0.08M_{\odot}$ and $0.1M_{\odot}$ disc as the inner regions are gravitationally stable. However in the $0.25M_{\odot}$ disc, as it is completely gravitationally unstable, the planets are unable to slow down until they reach the inner boundary of the disc. The dot in the left panel marks the location of a $1M_{\mathrm{Sat}}$ planet as it is undergoing a brief period of outward migration in a $0.08M_{\odot}$ disc (see \S\ref{sec:mockAlma} for details).}
    \label{fig:migration}
    \end{center}
\end{figure*}

\subsection{Disc setup}
The disc setup is identical to that in \cite{2020Rowther} \& \cite{2020bRowther}, and is summarised below. All discs are modelled using 2 million particles between $R_\mathrm{in} = 1$ and $R_\mathrm{out} = 25$ in code units where the fiducial disc has a disc-to-star mass ratio of 0.1. This is equivalent to a $0.1M_{\odot}$ disc around a $1M_{\odot}$ star. Sink particles \citep{1995Bate} are used to model both the central star and the planet. The accretion radius of the central star is set to be equal to the disc inner boundary, $R_\mathrm{in}$. The surface density profile $\Sigma$ is given by
\begin{equation}
\Sigma = \Sigma_{0}  \left ( \frac{R}{R_{0}} \right)^{-1} f_{s},
\end{equation}
where $\Sigma_{0}$ is the surface mass density at $R=R_{0}=1 $ and ${f_{s} = 1-\sqrt{R_\mathrm{in}/R}}$ is {the factor used to smooth the surface density at the inner boundary of the disc}. The initial temperature profile is expressed as a power law
\begin{equation}
T = T_{0} \left ( \frac{R}{R_{0}} \right)^{-0.5},
\end{equation}
where $T_{0}$ is set such that the disc aspect ratio ${H/R=0.05}$ at $R=R_{0}$. The energy equation is 
\begin{equation}
\label{eq:eng}
\frac{\mathrm{d}u}{\mathrm{d}t} = -\frac{P}{\rho} \left ( \nabla \cdot \vb*{v} \right) + \Lambda_{\mathrm{shock}} - \frac{\Lambda_{\mathrm{cool}}}{\rho}
\end{equation}
where we assume an adiabatic equation of state, and $u$ is the specific internal energy, $P$ is the pressure, $\rho$ is the density and $\vb*{v}$ is the velocity. The first term on the RHS is the $P\mathrm{d}V$ work, and $\Lambda_{\mathrm{shock}}$ is a heating term that is due to the artificial viscosity used to correctly deal with shock fronts. The final term  
\begin{equation}
\Lambda_{\mathrm{cool}} = \frac{\rho u}{t_{\mathrm{cool}}}
\end{equation}
controls the cooling in the disc. Here the cooling time is straightforwardly implemented to be proportional to the dynamical time by a factor of $\beta(R)$,
\begin{equation}
t_{\mathrm{cool}} = \beta(R)\Omega^{-1} = \beta_{0} \left( \frac{R}{R_{0}} \right)^{-2} \Omega^{-1},
\end{equation}
where $\Omega$ is the orbital frequency and we have varied $\beta$ with radius \citep{2020Rowther}. In addition to the disc mass, the strength of gravitational instability also depends on the cooling factor \citep{2009Cossins}. Hence, we set $\beta_0 = 5500$. The cooling factor is $\beta = 13.75$ and $8.8$ at the planet's location, $R_p = 20$ and at $R_\mathrm{out}$, respectively. This allows us to mimic a realistic self-gravitating disc that is only gravitationally unstable in the outer regions \citep{2005Rafikov, 2009Stamatellos, 2009Rice, 2009Clarke}, while being comparable to constant $\beta$ simulations in the outer regions.

\cite{2010Cullen} introduced an artificial viscosity switch that utilises the time derivative of the velocity divergence, which we use here to model shocks. The artificial viscosity parameter $\alpha_{\text{AV}}$ varies depending on the proximity to a shock. Close to the shock, it takes a maximum of $\alpha_{\mathrm{max}} = 1$, and a minimum of $\alpha_{\mathrm{min}} = 0$ far away. The artificial viscosity coefficient $\beta_{\text{AV}}$ is set to 2 (see \citealt{2018Price,2015Nealon}).

\subsection{The Suite of Simulations}

In our suite we consider three different disc masses and three planet masses. We begin by simulating three discs -- that differ only in their mass -- for 10 orbits at $R = R_\mathrm{out}$ to allow spiral structure in the disc to develop. The final snapshot from each of these three simulations are then used as the initial condition for four subsequent simulations; three of these containing embedded planets of differing masses and one without a planet. The latter is a control simulation to compare the impact a migrating giant planet has on the spiral structure in a gravitationally unstable disc.

To investigate the importance of the planet mass on suppressing spiral structures in a GI disc, we choose $1M_{\mathrm{Sat}}$, $1M_{\mathrm{Jup}}$, and a $3M_{\mathrm{Jup}}$ for our planet masses. Each planet is embedded (in code units) at $R_{p}=20$ with an accretion radius of 0.001, and the simulation is run for another 8 orbits, with a total simulation time of 18 orbits.
To investigate the importance of the disc mass on planet-disc interactions in a GI disc, for our disc masses we choose a less gravitationally unstable $(0.08M_{\odot})$ and a more gravitationally unstable $(0.25M_{\odot})$ disc in addition to our fiducial $(0.1M_{\odot})$ disc. Figure \ref{fig:InitialDisc} shows each of the three discs right before the planet is embedded.

\subsection{Post processing of simulations}

The raw synthetic continuum images at $1.3$mm are created using \textsc{mcfost} \citep{2006Pinte, 2009Pinte}. The simulations are scaled such that the disc size is $200$\textsc{au}, and initial ${R_{p}=160}$\textsc{au}. We use $10^{8}$ photon packets on a Voronoi tesselation where each \textsc{mcfost} cell corresponds to an individual SPH particle. The luminosity of the star is calculated assuming a mass of $1M_{\odot}$ and a 1Myr isochrone from \cite{2000Siess} which corresponds to a temperature of $T_{\star} = 4286$K. We assume that the dust is perfectly coupled to the gas and a constant dust-to-gas ratio of 0.01. As the Stokes numbers in all the simulations in this work are less than unity, this assumption is valid. Additionally, in a GI disc, the dust is stil able to couple quite well to the gas even at Stokes numbers of ${\sim}10$ \citep{2021Baehr}. The dust sizes vary between 0.3 and 1000 $\mu$m and are distributed across 100 different sizes with a power-law exponent of -3.5. We assume all dust grains are made of astronomical silicates, and are spherical and homogeneous. We compute the dust properties using Mie theory. The disc is assumed to be at a distance of 140pc.

Mock millimeter continuum observations are created using the ALMA Observation Support Tool \citep{2011Heywood}. We use an integration time of 60 minutes in the ALMA Cycle 9 C-7 configuration. We assume a bandwidth of 7.5GHz and a precipitable water vapour level of 0.913mm. \texttt{CLEAN} images are created using natural weights resulting in a beam size of $0.107{}'' \times 0.124{}''$, or equivalently $15.0$\textsc{au} $\times 17.3$\textsc{au} at 140pc.

In \cite{2020bRowther}, the kink caused by a $3M_{\mathrm{Jup}}$ planet in a $0.1M_{\odot}$ disc was not visible in the channel maps of the optically thick $^{12}$CO. Therefore in this work, the kinematics is investigated using just the optically thinner $^{13}$C$^{16}$O. Assuming a disc inclination and position angle of $40^{\circ}$, channel maps are generated for the ${J=3-2}$ transition using a velocity resolution of $0.1$ km/s. We choose this transition as it can be observed with a good compromise between observation time and signal-to-noise ratio. The abundance of $^{13}$C$^{16}$O is assumed to be a fraction $7\times 10^{-7}$ of the total disc mass (i.e. relative to $H_{2}$) respectively. Unless specified, we do not account for CO freeze-out at $T < 20$K, and photo-dissociation and photo-desorption in regions of high UV radiation (see Appendix B of \citealt{2018Pinte}).

\begin{figure*}
    \begin{center}
    \includegraphics[width=\textwidth]{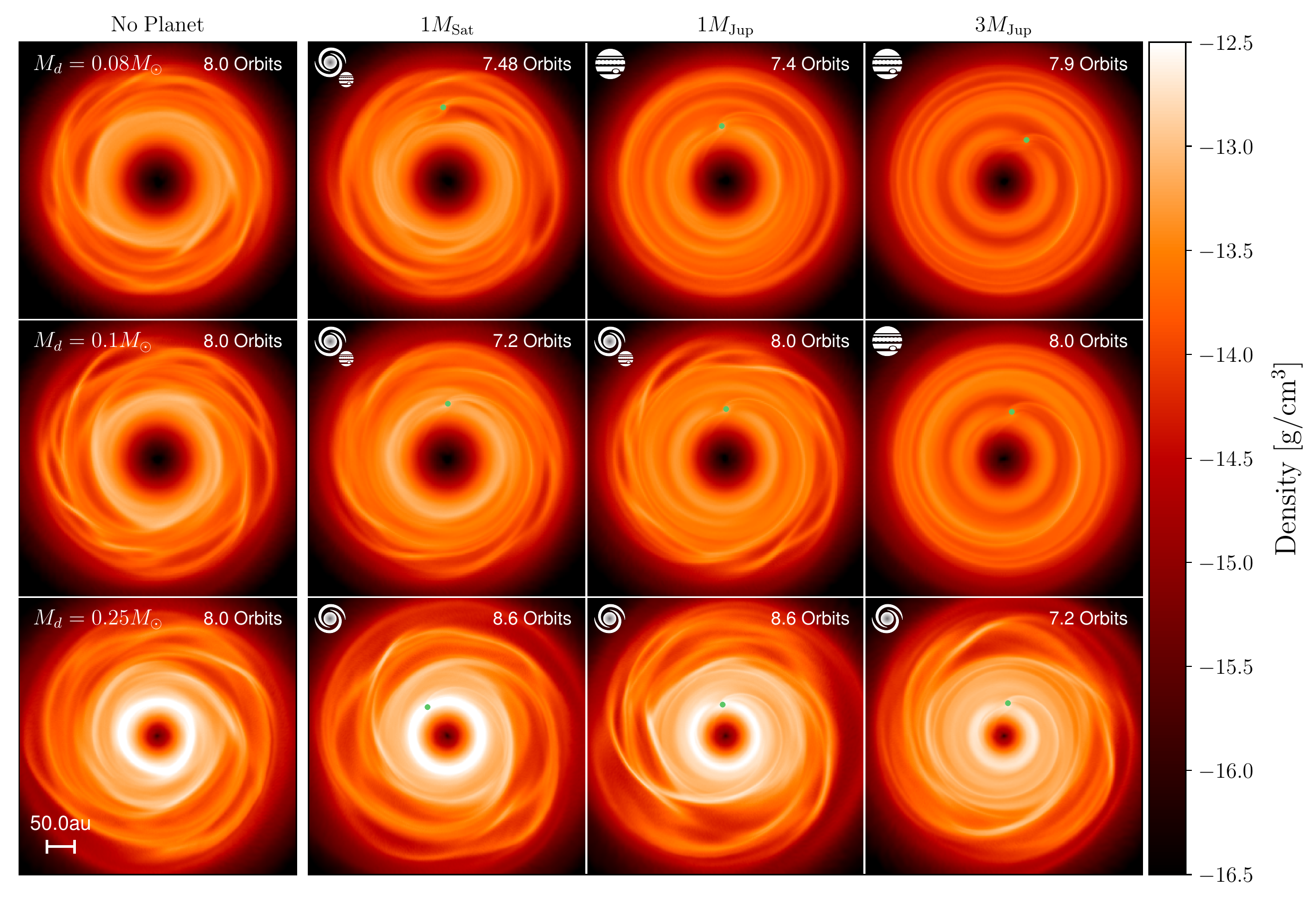}
    \caption[]{Density rendered plots of various disc and planet masses. From top to bottom, the disc masses are $0.08M_{\odot}$, $0.1M_{\odot}$, and $0.25M_{\odot}$. The left-most, detached column shows how the disc evolves without a planet. The rest from left to right contain a $1M_{\mathrm{Sat}}$, $1M_{\mathrm{Jup}}$, and a $3M_{\mathrm{Jup}}$ planet, shown by a green dot. These show that if a planet is massive enough, relative to the disc mass, it can completely suppress spiral structures due to GI, e.g. a $0.08M_{\odot}$ or $0.1M_{\odot}$  disc with a $3M_{\mathrm{Jup}}$ planet. This regime is highlighted by the \planet\ icon. If its mass is too low, the disc remains gravitationally unstable, e.g. with any planet in a $0.25M_{\odot}$ disc, highlighted by the \GI\ icon. The simulations highlighted by the \GIplanet\ icon show when the planet mass lies somewhere between those two extremes, spiral structures due to GI are weakened, e.g. a $0.1M_{\odot}$ disc with a $1M_{\mathrm{Jup}}$ planet.}
    \label{fig:density}
    \end{center}
\end{figure*}

\section{Results}
\label{sec:results}

Once the disc develops spiral structures, two different scenarios are followed. In the first the simulation is continued as normal to show how the disc would evolve in the absence of any planets. In the second, a planet is embedded. This is done for all combinations of a $1M_\mathrm{Sat}$, $1M_\mathrm{Jup}$, and a $3M_\mathrm{Jup}$ planet in a $0.08M_{\odot}$, $0.1M_{\odot}$, and a $0.25M_{\odot}$ disc, totalling 9 simulations with planets. In all cases, the planet starts in the outer gravitationally unstable parts of the disc and migrates rapidly inwards. As the planet migrates, the spiral wakes generated by the planet begin influencing the global disc properties \citep{2001Goodman,Rafikov2016,2020Ziampras}. The final fate of the migrating planets and their impact on the disc structure depend on both the planet's mass and the strength of gravitational instabilities  (i.e. the disc mass).

\subsection{Planet Migration}

In the $0.08M_{\odot}$ and $0.1M_{\odot}$ discs, the variable $\beta$ model results in a disc with a gravitationally stable inner region as shown in \cite{2020Rowther}. Whereas, the $0.25M_{\odot}$ disc is massive enough to still be gravitationally unstable everywhere despite the variable $\beta$ model. The migration of a $1M_\mathrm{Sat}$, $1M_\mathrm{Jup}$, and a $3M_\mathrm{Jup}$ planet is shown in Figure \ref{fig:migration} in each of the three discs. Fig \ref{fig:migration} demonstrates how the strength of gravitational instabilities, and thus the mass of the disc (increasing from left to right), determines the fate of the planet.

In the left and middle panels (the $0.08M_{\odot}$ and $0.1M_{\odot}$ disc), the migration of each planet slows down when it reaches the stable inner disc, but at different locations. The lighter the planet, the further out the planet slows down. After which, the planet continues to migrate inwards at a slower pace, consistent with the results in \cite{2020Rowther}. However, as the entire $0.25M_{\odot}$ disc is gravitationally unstable, the migration of the planets (right panel) is much more similar with simulations using a constant $\beta$ where the planet rapidly migrates inwards towards the inner boundary of the disc \citep{2011Baruteau,2015Malik,2020Rowther}. After which, migration ceases.

\begin{figure*}
    \begin{center}
    \includegraphics[width=\textwidth]{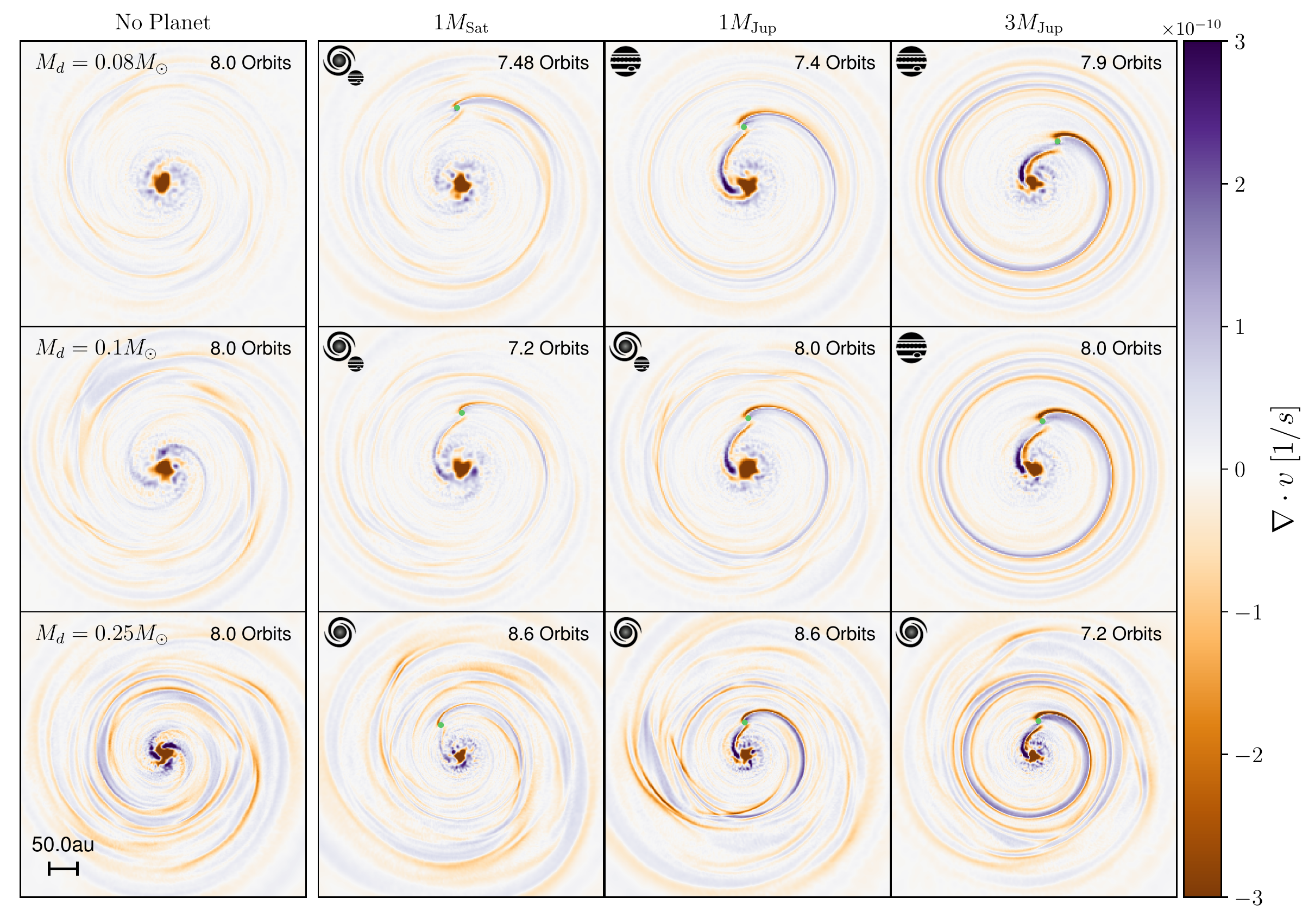}
    \caption[]{Velocity Divergence $\nabla \cdot \vb*{v}$ rendered plots of various disc and planet masses. The simulations are arranged as in Figure \ref{fig:density}. When the magnitude of $\nabla \cdot \vb*{v}$ due to the planet's spiral wake is greater relative to the $\nabla \cdot \vb*{v}$ due to GI, the evolution of the disc is driven by the planet resulting in a gravitationally stable disc with ring \& gap structure, e.g. a $0.08M_{\odot}$ or $0.1M_{\odot}$  disc with a $3M_{\mathrm{Jup}}$ planet. When the  magnitude of $\nabla \cdot \vb*{v}$ due to the planet and GI is comparable, GI structures are weakened, but not suppressed, e.g. a $0.1M_{\odot}$ disc with a $1M_{\mathrm{Jup}}$ planet. If the  magnitude of $\nabla \cdot \vb*{v}$ due to GI is dominant, then GI structures are unaffected, e.g. with any planet in a $0.25M_{\odot}$ disc.}
    \label{fig:divv}
    \end{center}
\end{figure*}

\subsection{Gravitational Instability vs Planet's Spiral Wakes}

Figures \ref{fig:density} and \ref{fig:divv} show the density and velocity divergence $\nabla \cdot \vb*{v}$ rendered plots for all the simulations in this work. In both figures, each row corresponds to a different disc mass. From top to bottom, the disc masses are $0.08M_{\odot}$, $0.1M_{\odot}$, and $0.25M_{\odot}$. Similarly, each column represents a different planet mass. The left-most detached column shows how the disc would have evolved in the absence of a planet. The rest from left to right contain a $1M_{\mathrm{Sat}}$, $1M_{\mathrm{Jup}}$, and a $3M_{\mathrm{Jup}}$ planet. The time at which the simulation is shown is chosen such that planets are on the North side of the disc, hence the times are not identical. Having the planet at roughly the same location makes for easier comparisons for mock observations in Figures \ref{fig:MockObs} and \ref{fig:channelMaps}, but the general results described here are applicable even at the same simulation time.
Similarly to \citet{2022Rowther}, we use the $\nabla \cdot \vb*{v}$ plots to understand how the spiral wakes of the planet are responsible for heating up the disc as from eq \ref{eq:eng}, changes in $\nabla \cdot \vb*{v}$ directly contribute to the energy equation, and thus the temperature of the disc.

From the various disc and planet masses explored in this work, we find that the planet's impact on the spiral structures due to GI can be quite different. For sufficiently massive discs, the gravitational instabilities are too strong to be suppressed by the planet. This is the case with the $0.25M_{\odot}$ disc in the bottom row of Fig \ref{fig:density}. Likewise, if the planet is massive enough, it can alter the evolution of the disc by heating it up enough to push it into the gravitationally stable regime, and thus suppressing spiral structures \citep{2020bRowther}. The suppression of GI is most apparent with the $3M_{\mathrm{Jup}}$ planet in the $0.08M_{\odot}$ and $0.1M_{\odot}$ discs. An in-between scenario is also seen where the planet mass is just large enough to weaken GI structures, but not enough to completely suppress it. The weakening of GI can be seen with both $1M_{\mathrm{Sat}}$ and $1M_{\mathrm{Jup}}$ planets in a $0.1M_{\odot}$ disc in the middle row of Fig \ref{fig:density}.

By expanding our parameter space beyond \cite{2020bRowther}, we can explore in more detail how the spiral wakes generated by the planet interacts with the gravitational instabilities present in the disc, seen in Figure 4. In general, the magnitude of $\nabla \cdot \vb*{v}$ due to GI is larger with increasing disc mass. This is expected as more massive discs are expected to be more gravitationally unstable. Similarly, $\nabla \cdot \vb*{v}$ due to the spiral wakes generated by the planet is also larger with increasing planet mass which is consistent with expectations of larger planets being more easily able to influence the disc's evolution \citep{2012Kley}. However, the impact of the planet on $\nabla \cdot \vb*{v}$ is broadly similar regardless of disc mass.

\begin{figure}
    \begin{center}
    \includegraphics[width=\linewidth]{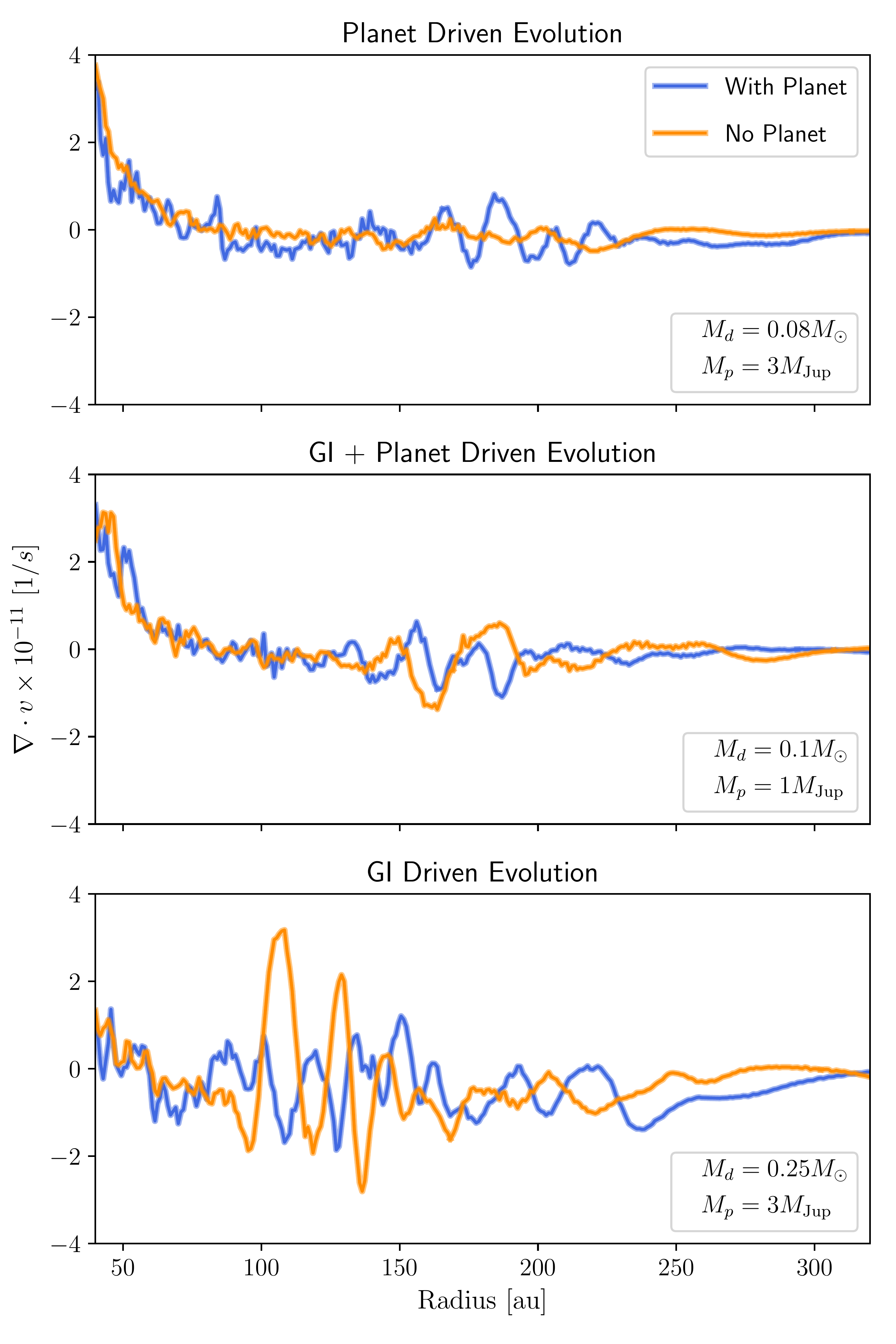}
    \caption[]{Azimuthally averaged $\nabla \cdot v$ plots showing the impact of a planet on representative simulations at the same times as in Figure \ref{fig:divv} that highlight the three scenarios that may occur. In this view, spiral structures in the disc are represented by radial variations in $\nabla \cdot v$.
    In the top panel, the spiral wakes generated by a $3M_{\mathrm{Jup}}$ planet in a $0.08M_{\odot}$ disc dominate over GI, thus controlling the evolution of the disc. In the middle panel, the magnitude of $\nabla \cdot \vb*{v}$ due to GI is comparable to a $1M_{\mathrm{Jup}}$ planet in a $0.1M_{\odot}$ disc. Hence, both the planet and GI play a role in the evolution of the disc. In the bottom panel, a $3M_{\mathrm{Jup}}$ planet in a $0.25M_{\odot}$ disc has a lesser impact on $\nabla \cdot \vb*{v}$ compared to GI. Thus the evolution of the disc is unchanged.}
    \label{fig:aziDivv}
    \end{center}
\end{figure}

\begin{figure}
    \begin{center}
    \includegraphics[width=\linewidth]{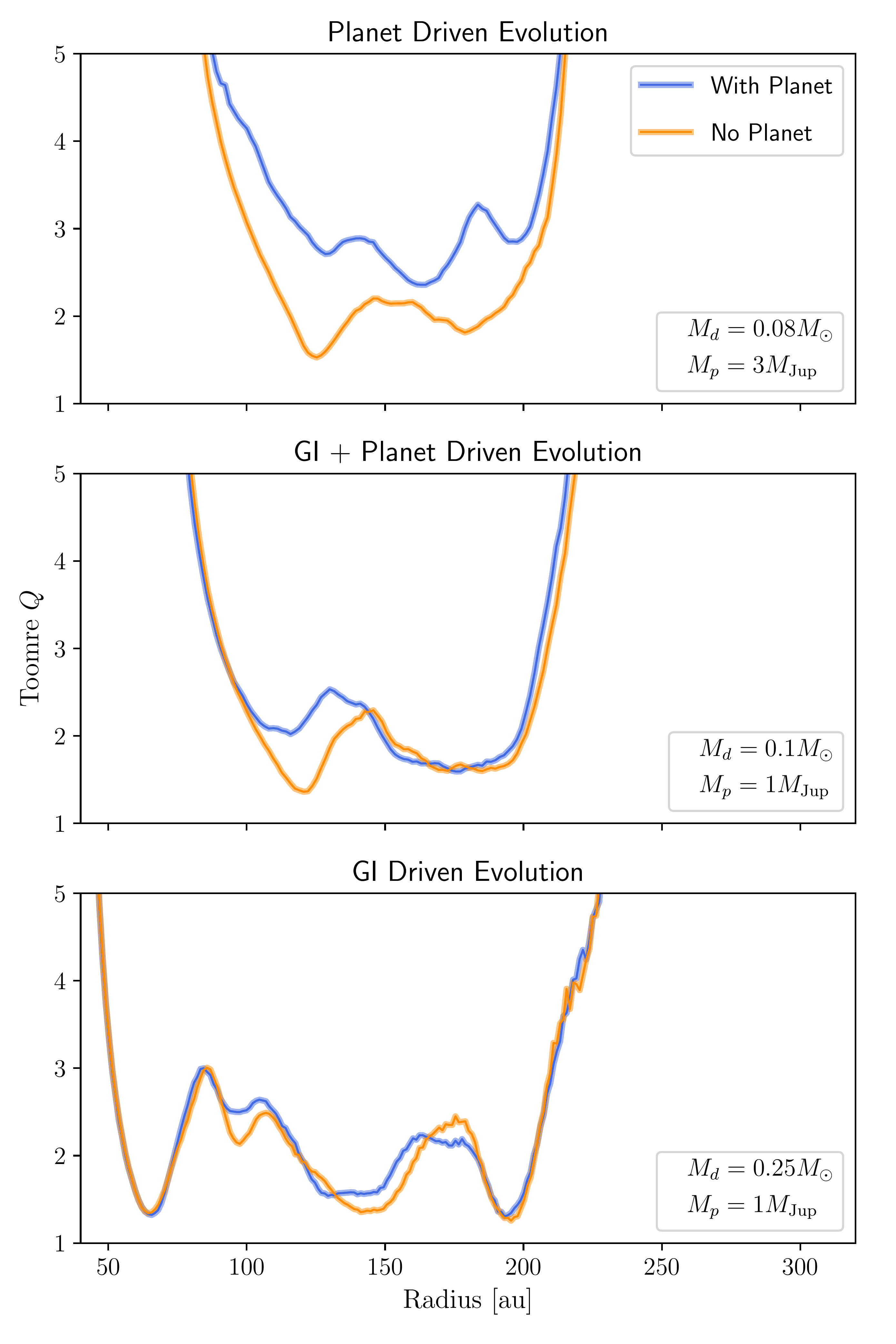}
    \caption[]{Azimuthally averaged Toomre $Q$ plots after two orbits showing the impact of a planet on three representative simulations that highlight the three scenarios that may occur. In the top panel, a $3M_{\mathrm{Jup}}$ planet in a $0.08M_{\odot}$ disc increases $Q$ globally, thus controlling the evolution of the disc and causing it to become gravitationally stable. In the middle panel, a $1M_{\mathrm{Jup}}$ planet in a $0.1M_{\odot}$ disc is unable to affect $Q$ in the outermost regions of the disc, but has a minor impact on the inner regions. Hence, both the planet and GI play a role in the evolution of the disc. In the bottom panel, a $1M_{\mathrm{Jup}}$ planet in a $0.25M_{\odot}$ disc has little impact on $Q$, thus the evolution of the disc is unchanged.}
    \label{fig:Q}
    \end{center}
\end{figure}

Figure \ref{fig:aziDivv} shows the azimuthally averaged $\nabla \cdot \vb*{v}$ of a subset of the simulations in Fig \ref{fig:divv}. In each panel the blue lines represent a simulation with a planet that is representative of one of the aforementioned scenarios, while the orange lines are for discs that continued to evolve without a planet. They highlight the increasing dominating effect of GI on the magnitude of  $\nabla \cdot \vb*{v}$ as the disc mass increases. Figures \ref{fig:divv} and \ref{fig:aziDivv} now reveal why none of the planets have a significant impact on the spiral structures due to GI in the most massive disc. The magnitude of $\nabla \cdot \vb*{v}$ due to GI is larger comparable to $\nabla \cdot \vb*{v}$ due to the planet's wake. As a result, the planet does not contribute much to the energy equation (Eq. \ref{eq:eng}) and the disc remains gravitationally unstable. Only the $3M_{\mathrm{Jup}}$ planet has a minor impact on the disc structure. However, the planet's wake is restricted to the inner disc. Thus, the outer disc remains mostly unaffected. Similarly, Figs \ref{fig:divv} and \ref{fig:aziDivv} also reveal that gravitational instabilities are completely suppressed when $\nabla \cdot \vb*{v}$ due to the planet's wake is dominant, as shown in the simulations with a $3M_{\mathrm{Jup}}$ planet in a $0.08M_{\odot}$ and $0.1M_{\odot}$ discs, and a $1M_{\mathrm{Jup}}$ planet in the $0.08M_{\odot}$ disc. In the in-between scenario where GI is weakened, this occurs when both GI and the planet have a comparable influence on $\nabla \cdot \vb*{v}$ as seen when there is a $1M_{\mathrm{Sat}}$ or $1M_{\mathrm{Jup}}$ planet in the $0.1M_{\odot}$ disc, and a $1M_{\mathrm{Sat}}$ planet in the $0.08M_{\odot}$ disc.

\begin{figure*}
    \begin{center}
    \includegraphics[width=0.9\textwidth]{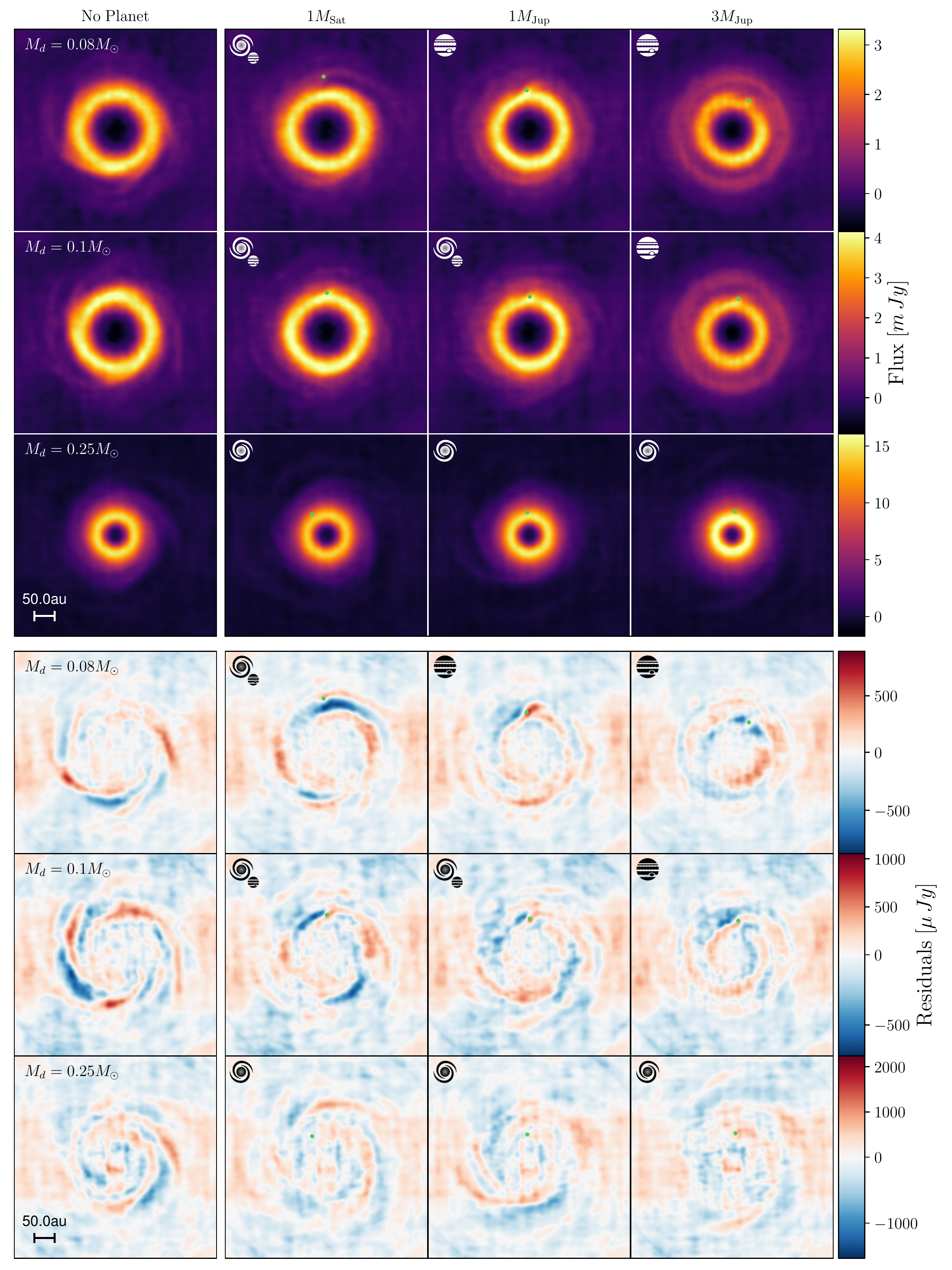}
    \caption[]{Upper panel: mock observations at 1.3mm using the C-7 configuration with an integration time of 60 minutes. Lower panel: the residuals of subtracting the mock observations from an axisymmetric model which highlights non-axisymmetric features. The simulations are arranged as in Figure \ref{fig:density}. The planet (represented by the green dot) can completely suppress spiral structures if it's massive enough, or have no affect on GI if its mass is too low, or weaken GI and making it harder to observe if the planet mass lies between the two extremes. The observability of spiral structures does however depend on the configuration and integration time used.}
    \label{fig:MockObs}
    \end{center}
\end{figure*}

The Toomre Q parameter gives a measure of how gravitationally unstable a disc is, and is defined by \citep{1964Toomre}
\begin{equation}
    Q = \frac{c_s \Omega}{\pi G \Sigma}.
\end{equation}
To best highlight the three scenarios that may occur, the azimuthally averaged Toomre Q plots in Figure \ref{fig:Q} are shown two orbits after the planet has been embedded for three representative simulations. To summarise;
\begin{description}
    \item[i) Planet Driven Evolution \planet\ --] The spiral wakes generated by a $3M_{\mathrm{Jup}}$ planet in a $0.1M_{\odot}$ or $0.08M_{\odot}$ have a large effect on $\nabla \cdot \vb*{v}$ causing the disc to heat up as seen by the global increase in the Q profile. Thus, GI structures are completely suppressed as the disc is pushed into the gravitationally stable regime.
    \setlength\itemsep{0.25em} 
    \item[ii) GI + Planet Driven Evolution \GIplanet\ --] The spiral wakes of a $1M_{\mathrm{Jup}}$ planet in a $0.1M_{\odot}$ disc are less extended compared to the $3M_{\mathrm{Jup}}$ planet. As the spiral wakes do not extend far out, the  the Q profile of outer parts of the disc remain in the gravitationally unstable regime. Thus the simulations still show GI structures in the outer regions.
    \item[iii) GI Driven Evolution \GI\ -- ] The spiral wakes generated by any of the three planets in a $0.25M_{\odot}$ disc are unable to have a major impact on $\nabla \cdot \vb*{v}$, and thus they have very little effect on the Q profile. Hence the disc remains just as gravitationally unstable as if it had evolved without a planet.
\end{description}

\begin{figure*}
    \begin{center}
    \includegraphics[width=\textwidth]{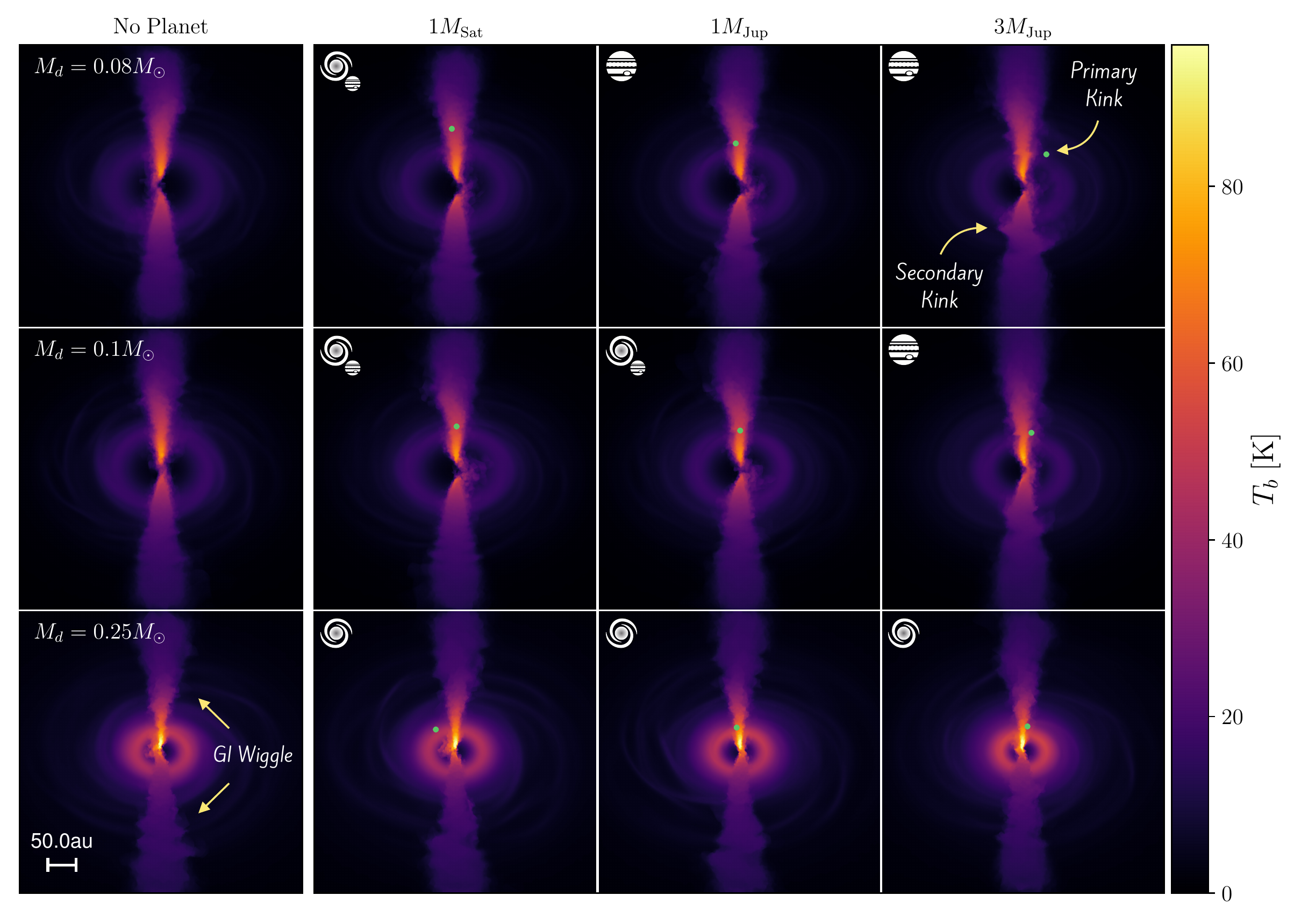}
    \caption[]{Synthetic channel maps (J = 3 -- 2 transitions) for $^{13}$C$^{16}$O at $\Delta v=0$ km s$^{-1}$ from the systemic velocity of the
    disc. The simulations are arranged as in Figure \ref{fig:density}. When the planet is able to open up a gap in the disc, a primary kink is seen at the planet's location along with a secondary kink on the opposite side due to the planet's spiral wake. GI wiggles are only clearly seen for the $0.25M_{\odot}$ disc.}
    \label{fig:channelMaps}
    \end{center}
\end{figure*}

\subsection{Mock ALMA observations}
\label{sec:mockAlma}

The top half of Figure \ref{fig:MockObs} shows mock observations for all simulations in this work at the same times as shown in Figures \ref{fig:density} and \ref{fig:divv} at 1.3mm with an integration time of 60 minutes using the C-7 configuration. The bottom half highlights non-axisymmetric features by plotting the residuals of subtracting the mock observations from an axisymmetric model. From top to bottom, the disc masses are $0.08M_{\odot}$, $0.1M_{\odot}$, and $0.25M_{\odot}$. The left-most detached column shows how the disc would be observed in the absence of a planet. The rest from left to right show how the disc would appear if it contained a $1M_{\mathrm{Sat}}$, $1M_{\mathrm{Jup}}$, and a $3M_{\mathrm{Jup}}$ planet.

From Figure \ref{fig:density}, in a $0.25M_{\odot}$ disc the planets do not have much, if any, impact on gravitational instabilities in the disc. This is reflected in the mock observations in Figure \ref{fig:MockObs}, where both the flux and residual plots are similar regardless of whether a planet is present in the disc. 

In contrast, Figure \ref{fig:density} shows that the $0.08M_{\odot}$ and $0.1M_{\odot}$ discs are much more affected by the presence of a planet. If the planet is massive enough and drives the disc's evolution, as is the case with a $3M_{\mathrm{Jup}}$ planet in a $0.08M_{\odot}$ and $0.1M_{\odot}$ discs, ring \& gap structure is seen in the mock observations. The weakening of GI by the lower mass planets is also reflected in the mock observations where evidence of GI in the form of large-scale spiral structures is far less clear. The lack of obvious GI structures is highlighted by the plots in the left-most detached column which show that in the absence of a planet, evidence of GI is apparent in the mock observations, and especially the residuals in Fig \ref{fig:MockObs}.

The planet leaves a few hints of its presence in Fig \ref{fig:MockObs}. For clarity, this will be described using the mock observations of a $3M_{\mathrm{Jup}}$ planet in either the $0.08M_{\odot}$ or $0.1M_{\odot}$ disc as the planet dominates over GI, where all the residual features can be confidently attributed to the planet. The largest of the features in the residuals show that the rings are asymmetric in brightness, highlighted by the red regions in the vicinity of the planet. Comparison with Fig \ref{fig:density} reveals that this asymmetry can be attributed to overdense regions resulting from the spiral wakes of the planet. As mentioned earlier, the gravitationally stable inner disc slows down planet migration in the $0.08M_{\odot}$ or $0.1M_{\odot}$ discs. The inward migration is due to a large negative coorbital torque resulting from an underdense region in front of the planet \citep{2020Rowther,2015Malik,2011Baruteau}, which is revealed in the residuals as a blue (underdense) region in front of the planet. The opposite is seen for an outwardly migrating $1M_{\mathrm{Sat}}$ planet in a $0.08M_{\odot}$ disc where the blue (underdense) region is behind the planet. The time at which the mock observation is created is marked in the migration track in Figure \ref{fig:migration} highlighting the outward migration. The planet also appears as a bright spot in the mock continuum data. It is likely that this bright spot is artificially enhanced by our choice of accretion radius, see \S \ref{sec:disc_obs}.

\subsection{Impact on the Kinematics}
\label{sec:kinematics}

The gas channel maps can also be used to detect GI or the presence of giant planets \citep{2019Pinte,2020Pinte,2020Hall}. Figure \ref{fig:channelMaps} shows the synthetic channel maps at $\Delta v = v_{\mathrm{obs}} - v_{\mathrm{systemic}} = 0$ km/s for all the simulations in this work. In real observations, the channel maps would be affected by CO freeze-out, and photodissociation and photodesorption in regions of high UV radiation. However, the channel maps in Fig \ref{fig:channelMaps} are created excluding these effects for easier comparison with \cite{2020Hall} and \cite{2022Terry}. Additionally, these have not been convolved with a beam size. Both are discounted for clarity in comparing the kinematic signatures of GI and a planet.

\subsubsection{Primary Kink}

In the case of a planet, its presence can be detected by a localised kink near the location of the planet \citep{2019Pinte,2020Pinte}. This is most apparent for a $3M_{\mathrm{Jup}}$ planet in a $0.1M_{\odot}$ or $0.08M_{\odot}$ disc where a kink is visible near the vicinity of the planet and is not seen at all radii. A weaker kink is also visible for a $1M_{\mathrm{Jup}}$ planet in both the $0.1M_{\odot}$ and $0.08M_{\odot}$ discs, but this is likely to be detectable only in synthetic observations which are not affected by noise and not smoothed by the beam size.  However, if the planet's influence does not dominate over GI, the planet is undetectable from the kinematics as evident by the lack of any localised kink in the $0.25M_{\odot}$ disc.

\subsubsection{Secondary Kink}

The presence of a planet in the kinematics can be detected by any extended features caused by the spiral wake of the planet. In Fig \ref{fig:channelMaps}, this extended feature is visible as a secondary kink on the opposite side of the planet. The imprint of the planet's spiral wake is most easily seen for a $1$ and $3M_{\mathrm{Jup}}$ in a $0.08M_{\odot}$ disc, and a $3M_{\mathrm{Jup}}$ in a $0.1M_{\odot}$ disc. The imprint left behind by the wake of the planet as seen in these simulations have also been detected in the disc around HD 163296 \citep{2021Calcino}.

\subsubsection{GI Wiggle}

In the case of GI, its presence can be detected in the form of \textit{`GI wiggles'}  \citep{2020Hall}. The global nature of GI results in these wiggles being non-localised appearing at all velocities and radii in the disc. In Fig \ref{fig:channelMaps}, a clear GI wiggle is only seen for a $0.25M_{\odot}$ disc where strong gravitational instabilities are present. The amplitude of the wiggle is larger further out in the disc. Since $\beta$ decreases with radius, this behaviour is consistent with \cite{2021Longarini} which showed the amplitude of the wiggle is stronger for lower values of $\beta$. As mentioned earlier, when GI is stronger compared to the planet's spiral wake, the disc structure is unaffected. This is reflected in the kinematics by the lack of any features associated with the planet visible in the $0.25M_{\odot}$ disc.

Despite being gravitationally unstable and harbouring large scale spiral features, the $0.08M_{\odot}$ and $0.1M_{\odot}$ discs only show tentative GI wiggles, which is likely to be hidden when observational effects (such as a beam size) are included. However, this is expected from previous work which showed GI wiggles are weaker for lower disc masses \citep{2021Longarini,2022Terry}.

\subsubsection{Including CO Freeze-out, Photodissociation and Photodesorption}
\label{sec:kinkObs}

\begin{figure}
    \begin{center}
    \includegraphics[width=\linewidth]{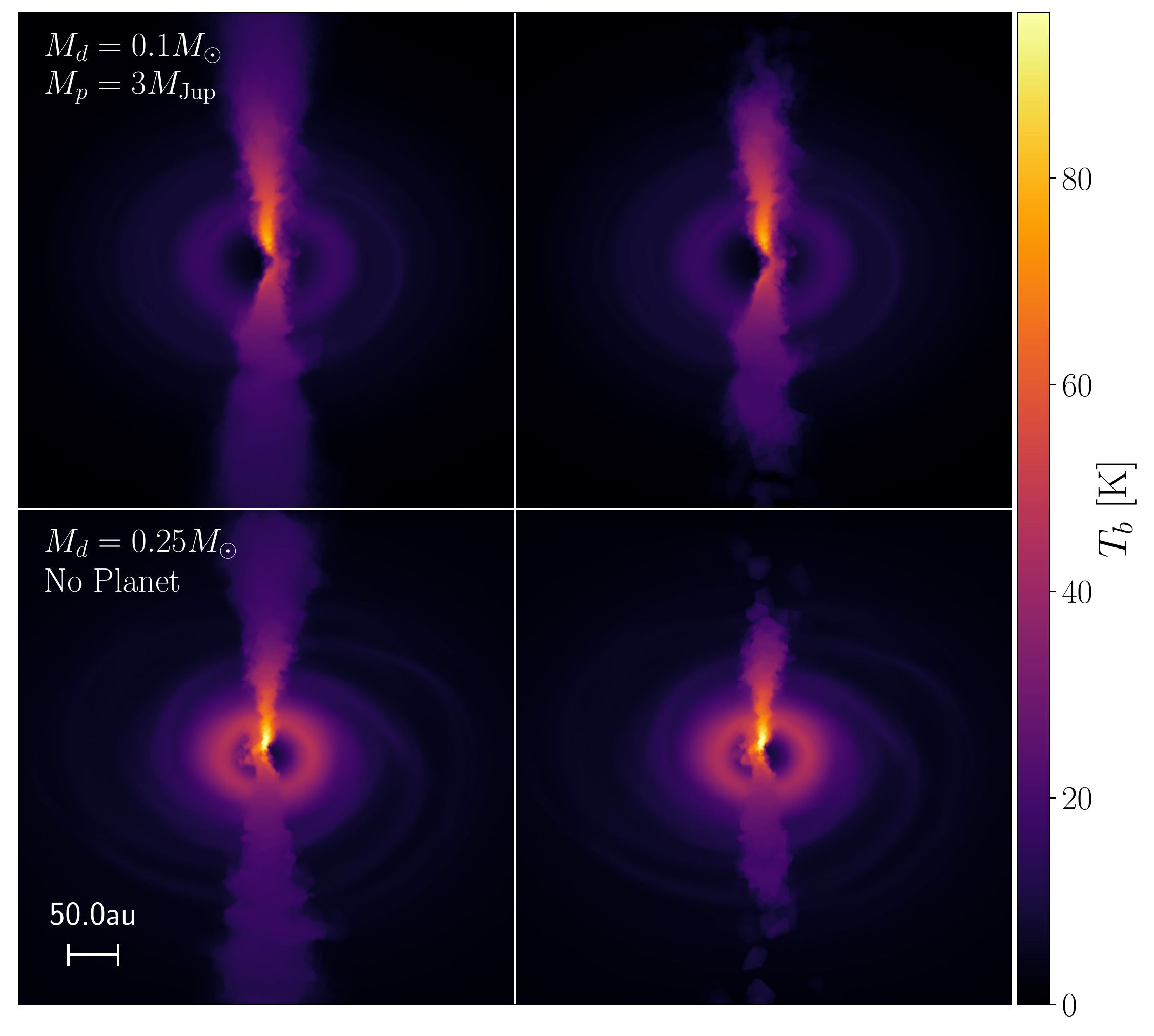}
    \caption[]{Synthetic channel maps (J = 3 -- 2 transitions) for $^{13}$C$^{16}$O at $\Delta v=0$ km s$^{-1}$ from the systemic velocity of the
    disc. The top panels are for a $3M_{\mathrm{Jup}}$ planet in a $0.1M_{\odot}$ disc, while the bottom panels are for a $0.25M_{\odot}$ disc without a planet. The right panels includes CO freeze-out below 20K, and photodissociation and photodesorption in regions of high UV radiation, while the left panels do not. Inclusion of these observational effects do not affect the visibility of the kinematic features of the planet. However, as GI is strongest in the outer colder regions of the disc, they are more affected and are less visible.}
    \label{fig:compFreezeout}
    \end{center}
\end{figure}

To determine if the kinematic features due to a planet or GI are impacted by observational effects, Figure \ref{fig:compFreezeout} compares the channel maps with (right panels) and without (left panels) the inclusion of CO freeze-out below 20K, and photodissociation and photodesorption in regions of high UV radiation. The comparison is only presented for two simulations; a $3M_{\mathrm{Jup}}$ planet in a $0.1M_{\odot}$ disc and a $0.25M_{\odot}$ disc without a planet. The former is chosen to investigate the observability of the kinematic signatures due to a planet which shows that both the primary and secondary kink are unaffected. The latter is chosen to determine observability of GI structures in the channel maps which show that the GI wiggles are far less visible due to the inclusion of the aforementioned observation effects.

\section{Discussion}
\label{sec:disc}

\subsection{Observational Implications}

Simulations of star cluster formation \citep{2018Bate} show that discs are expected to be massive and potentially gravitationally unstable in the earliest stages of their lives. A characteristic feature of such massive discs are their large-scale spiral arms. Despite the increasing number of resolved protoplanetary discs, these type of discs remain quite rare in observations \citep{2016Perez,2018Huang}. Additionally, more complex methods have measured disc masses to be higher than previous estimates using a constant fixed dust-to-gas mass ratio to infer the total mass of the disc. Using the rare optically thinner $^{13}$C$^{17}$O, \cite{2020Booth} measured the mass of HL Tau to be large enough to enter the gravitationally unstable regime. Utilising the aerodynamic properties of dust grains \cite{2019Powell} measured new disc masses to be 9-27\% of their stellar hosts. Such high masses push the discs close to the gravitationally unstable regime where the disc would be expected to show evidence of gravitational instabilities, but instead they show axisymmetric ring \& gap structures. Our results show that weak and moderate gravitational instabilities can be suppressed by planet-disc interactions, thus highlighting that massive discs cannot be simply dismissed by the lack of large-scale spiral structures in the disc midplane. \cite{2021Veronesi} showed that the self-gravity of the Elias 2-27 protoplanetary disc can have a noticeable impact on its rotation curve. Hence, our results provide additional motivation for obtaining high resolution molecular data for axisymmetric discs so that dynamical mass estimates can be calculated to determine if their spiral structures due to GI are hidden by planets or other processes.

\subsection{Observability of Continuum and Kinematic Features}
\label{sec:disc_obs}

The results presented here have made a few assumptions and simplifications which affect the observability of the continuum and kinematic substructures described in \S\ref{sec:results}. 

The bright spot at the planet's location in our synthetic observations is most likely to be impacted due to the simplification of using an accretion radius of 0.001 code units. We choose a small accretion radius to ensure the planet does not accrete too much material. If the planet was allowed to freely accrete, it would quickly reach the threshold required to open up a gap at which point all the planets would behave identically, which defeats the purpose of investigating the impact of different planet masses. The consequence of such a tiny accretion radius is a lot of gas swirling around the planet. Although a circumplanetary disc is an expected physical feature, resolving it is not the purpose of this study and it is unlikely that we do so in these simulations.

In this work, the dust has been assumed to be perfectly coupled to the gas. Consequently, any asymmetric structure due to the planet present in the simulation is also evident in the mock continuum observation. If the dust is decoupled from the gas, the asymmetric features are less likely to be observable \citep{2015Dipierro} as the dust will be able to form substructures that do not trace the gas substructures perfectly. However, given the high masses of the disc involved the Stokes numbers of millimeter sized grain is smaller than one and thus are more likely to be coupled to the gas. Nevertheless, simulations with a gas and dust mixture are required to test the validity of this assumption further. Additionally, the observability of these substructures also depend on ALMA configuration and integration time used. Recently, \cite{2022Speedie} also found that planet driven spirals could be observable with ALMA with longer integration times. This is also true for the observability of GI structures when they are weakened by the planet like with a $1M_{\mathrm{Sat}}$ or $1M_{\mathrm{Jup}}$ planet in the $0.1M_{\odot}$ disc.

The cavity seen in all simulations is due to the inner-most regions of the disc not being resolved. Consequently, the inner boundary of the disc shows up as a ring in the continuum. Higher resolution and more realistic thermodynamics could both resolve these issues.

In \S\ref{sec:kinkObs} it was shown that the inclusion of CO freeze-out, and photodissociation and photodesorption had more of an effect on the kinematic signatures of GI rather than a planet. Gravitational instabilities are expected to be strongest in the outer colder regions of the disc. Hence, their signature GI wiggle is more likely to be affected by CO freeze-out. This is seen in the bottom right panel of Fig \ref{fig:compFreezeout} by the lack of emission (and thus wiggles) in the outer parts of the disc. However, as the planet is located in the inner warmer parts of the disc, the kinematic features associated with the planet are still visible with the extra physics included.

\subsection{Cooling prescription}

A caveat to the simulations presented in this study is the cooling prescription used. Although it is able to mimic a realistic gravitationally unstable disc, i.e. one that only has GI structures in the outer regions of the disc, it does not account for the evolution of the disc. Using radiative transfer simulations, \cite{2018Mercer} calculated an effective $\beta$ which was found to vary both spatially and with time. While this does not change our  conclusions qualitatively, the long term evolution of planet-disc interactions in a GI disc requires studies with more realistic thermodynamics.

\section{Conclusions}
\label{sec:conc}

We perform 3D SPH simulations to investigate the impact of planet-disc interactions on the structure of a gravitationally unstable disc. Our work shows that the more gravitationally unstable a disc is, the harder it is for a planet to influence the disc's evolution. However, the planet can affect the disc's evolution when the strength of the planet's spiral wake is comparable to or larger than gravitational instabilities. When the planet is massive enough, it is able to open up a gap in the disc.

The mock ALMA observations of the continuum show that the observability of GI structures is either diminished or completely suppressed when the planet is massive enough to impact the disc structure. The analysis of the kinematics reveals that if either the kinematic signature of a planet or GI is present, then the other is not detected. If the planet drives the disc's evolution, then the planet's imprint is visible in the kinematics. The GI wiggle is only clearly visible for the highest mass disc, and as the planet is not massive enough to impact the disc structure, the GI wiggles are not affected by the presence of the planet.

Our results show that it is possible for high mass discs to appear axisymmetric, and lacking large-scale spiral structures in the presence of a giant planet. The implications on the origins of disc substructures, planet formation, and disc evolution necessitate these findings to be taken into account when considering the many possible discrepancies between the disc structure and inferred mass in observations.

\section*{Data Availability}

The data from the simulations used to create all plots in this article is available on reasonable request to the corresponding author. The software used to create and visualise the simulations, \textsc{Phantom} and \textsc{Splash} respectively, are publicly available at \url{https://github.com/danieljprice/phantom} and 
\url{https://github.com/danieljprice/splash}.

\section*{Acknowledgements}

We thank the anonymous referee for their useful comments which benefited this work. S.R. acknowledges support from a Royal Society Enhancement Award, and funding from the Science \& Technology Facilities Council (STFC) through Consolidated Grant ST/W000857/1. F.M. acknowledges support from the Royal Society Dorothy Hodgkin Fellowship. R.N. acknowledges support from UKRI/EPSRC through a Stephen Hawking Fellowship (EP/T017287/1). This work was performed using Orac, the HPC cluster at the University of Warwick.



\bibliographystyle{mnras}
\bibliography{PlanetDiscInteractions} 








\bsp	
\label{lastpage}
\end{document}